\documentclass[11pt,a4paper]{article} 
\pdfoutput=1
\usepackage{jcappub}
\usepackage{graphicx}
\usepackage[lofdepth,lotdepth]{subfig}
\usepackage{dcolumn}
\usepackage{bm}
\usepackage{epsfig}
\usepackage{color}
\usepackage{rotating}
\usepackage{psfrag}

\title{Constraining Sterile Neutrinos with AMANDA and IceCube Atmospheric Neutrino Data}

\author[a,b]{Arman Esmaili}
\author[c]{, Francis Halzen}
\author[a,d]{, O. L. G. Peres}
\emailAdd{aesmaili@ifi.unicamp.br}
\emailAdd{halzen@icecube.wisc.edu}
\emailAdd{orlando@ifi.unicamp.br}
\affiliation[a]{Instituto de Fisica Gleb Wataghin - UNICAMP, 13083-859, Campinas, SP, Brazil}
\affiliation[b]{Kavli IPMU (WPI), The University of Tokyo, Kashiwa, Chiba 277-8583, Japan}
\affiliation[c]{Wisconsin IceCube Particle Astrophysics Center and Department of Physics, University of Wisconsin, Madison, WI 53706, USA}
\affiliation[d]{Abdus Salam International Centre for Theoretical Physics, ICTP, I-34010, Trieste, Italy}

\abstract{We demonstrate that atmospheric neutrino data accumulated with the AMANDA and the partially deployed IceCube experiments constrain the allowed parameter space for a hypothesized fourth sterile neutrino beyond the reach of a combined analysis of all other experiments, for $\Delta m^2_{41}\lesssim1\,{\rm eV}^2$. Although the IceCube data wins the statistics in the analysis, the advantage of a combined analysis of AMANDA and IceCube data is the partial remedy of yet unknown instrumental systematic uncertainties. We also illustrate the sensitivity of the completed IceCube detector, that is now taking data, to the parameter space of 3+1 model.}

\begin{document}

\maketitle

\section{\label{sec:intro}Introduction}

Data from atmospheric, solar, accelerator and reactor neutrino oscillation experiments have firmly established the pattern of oscillation of the three flavors of neutrinos~\cite{GonzalezGarcia:2007ib}. The origin of oscillation is the non-vanishing masses and non-trivial mixing of the neutrino states. The 3$\nu$ scheme of the neutrino sector of Standard Model has established the existence of two mass-squared differences $\Delta m_{\rm sol}^2\equiv \Delta m_{21}^2\simeq 8\times 10^{-5}\,{\rm eV}^2$ and $\Delta m_{\rm atm}^2=|\Delta m_{31}^2|\simeq |\Delta m_{32}^2|\simeq 2\times 10^{-3}\,{\rm eV}^2$ related, respectively, to solar and atmospheric neutrino flavor oscillations ($\Delta m_{ij}^2\equiv m_i^2-m_j^2$). The three mixing angles in the $3\nu$ scheme have also been measured: from solar~\cite{Ahmad:2002jz} and KamLand~\cite{Abe:2008aa} data $\sin^2\theta_{12}\simeq 0.3$; from atmospheric~\cite{Fukuda:1998mi} and MINOS~\cite{Michael:2006rx} data $\sin^2\theta_{23}\simeq0.5$. Recently the Double-CHOOZ~\cite{Abe:2011fz}, RENO~\cite{Ahn:2012nd} and Daya-Bay~\cite{An:2012eh} experiments established a non-zero value for the last mixing angle $\sin^2 \theta_{13}\simeq0.1$.    

Over the years a variety of  experimental results have challenged the $3\nu$ framework, although none with completely convincing statistics. The simplest extension of the $3\nu$ framework to accommodate these anomalies is the so-called $3+1$ scheme which assumes the presence of a fourth, mostly sterile, neutrino state with mass $m_4$. The first hint came from the LSND experiment in a search for the appearance $\bar{\nu}_e$ in a $\bar{\nu}_\mu$ beam over a base-line $L\simeq30$~m and neutrino energy $20~{\rm MeV}\leq E_\nu\leq 200~{\rm MeV}$~\cite{Aguilar:2001ty}. The possible observation of $\bar{\nu}_\mu\to\bar{\nu}_e$ by LSND can be accommodated by intorducing a new mass-squared difference $\Delta m_{41}^2\simeq 0.2-10~{\rm eV}^2$. The MiniBooNE experiment with base-line $L\simeq 540$~m and neutrino energy $200~{\rm MeV}\leq E_\nu\leq 1.25~{\rm GeV}$ and, therefore, the same $L/E_\nu$ as LSND, also has seen excess at 3.8$\sigma$ level in both $\bar{\nu}_\mu\to\bar{\nu}_e$ and $\nu_\mu\to\nu_e$ search channels~\cite{AguilarArevalo:2012va}.

Recently the re-evaluation of the $\bar{\nu}_e$ flux from reactors resulted in a 3.5\% increase in the expected flux~\cite{Mueller:2011nm,Huber:2011wv}. With this new reactor $\bar{\nu}_e$ flux, the combined results of all short base-line reactor neutrino experiments show a $3\sigma$ deficit in the observed versus predicted number of events, the so-called ``reactor anomaly"~\cite{Mention:2011rk}. It can also be interpreted as evidence for a new mass eigenstate with $\Delta m_{41}^2\simeq 1 \,{\rm eV}^2$. The observed $2.7\sigma$ deficit in the expected number of events in the calibration of the solar neutrino experiments GALLEX and SAGE, the so-called ``gallium anomaly"~\cite{Giunti:2010zu}, also hints on the presence of a new mass-squared difference of the same order as reactor anomaly. The combined reactor and gallium anomalies exclude the no-oscillation hypothesis at $3.6\sigma$ confidence level~\cite{Mention:2011rk}.

A hint on the presence of extra relativistic degrees of freedom also emerged from cosmology; the so-called ``dark radiation"~\cite{Hamann:2010pw,GonzalezGarcia:2010un,Archidiacono:2011gq} (also see~\cite{Abazajian:2012ys} and references therein). Although the significance of these hints depends on the data sample used in the analysis and on the assumptions regarding the cosmological models, various analyses favor the presence of light sterile neutrinos at $\sim2\sigma$ level with mass-squared difference $\Delta m_{41}^2\sim 0.1 \,{\rm eV}^2$.

While not compelling at the moment, it is imperative to test the evidence and one opportunity is provided by investigating the effects of light sterile neutrinos on the atmospheric neutrino flux. The partial propagation of atmospheric neutrinos through the Earth in a sterile state significantly modifies by the matter effects. With a mass-squared difference $\Delta m_{41}^2\sim 1 \,{\rm eV}^2$ matter effect induces a MSW active-sterile resonance flavor conversion of the $\bar{\nu}_\mu$ ($\nu_\mu$ for $\Delta m_{41}^2<0$ ) at the energies $\sim$~TeV~\cite{Nunokawa:2003ep,Choubey:2007ji}. The resonance leads to a distortion of the observed zenith angle distribution of atmospheric neutrino events. The construction of the km$^3$-scale neutrino detector IceCube at the South Pole, sensitive to neutrinos in this energy range, has opened a window to search for the telltale distortions in the spectrum of atmospheric neutrinos. Specifically, atmospheric neutrino data taken with IceCube-40~\cite{Abbasi:2010ie} ({\it i.e.} the half-completed IceCube-86 detector), have raised the possibility of obtaining improved fits to the zenith angle distribution within the 3+1 scheme, or, alternatively, using the measurement to constrain its parameter space~\cite{Razzaque:2012tp,Razzaque:2011ab,Barger:2011rc,Halzen:2011yq}.

In this paper we revisit the information that IceCube can provide on the possible existence of sterile neutrinos. Key is that our analysis combines the atmospheric data obtained by AMANDA-II over 6 years~\cite{Abbasi:2009nfa}  with those of IceCube-40~\cite{Abbasi:2010ie}. Analyses like this are at best illustrative because of the role of systematic errors which have to be evaluated by the experiments. The main systematic uncertainties can be divided in two categories: the physics uncertainties (such as the kaon to pion ratio and the spectral slope of the primary cosmic rays which produce the atmospheric neutrino flux) and instrumental uncertainties (originating from the geometry of detector, the angular sensitivity of the digital optical modules in the detector and the properties of the ice). The instrumental uncertainties are very different for the IceCube-40 and AMANDA-II experiments due to their different geometries and different depths. Although, due to the huge statistics of IceCube-40 data with respect to AMANDA-II, the main sensitivity of our analysis results from IceCube-40, combining data from two different experiments may partially remedy the problem of estimating instrumental systematic uncertainties. Our main conclusion is that the data do not support the existence of a fourth neutrino in the eV and sub-eV mass range and, on the contrary, yield limits on its parameter space that are stronger than those obtained from a combined analysis of all the other data~\cite{Giunti:2011cp}, for $\Delta m^2_{41}\lesssim1\,{\rm eV}^2$. We also present an estimate of the reach of the completed IceCube detector now taking data. Although the final sensitivities require an analysis that can only be performed by the individual experiments, we are confident that the relative sensitivities, discussed here, are reliably estimated.

After introducing the oscillation pattern in the presence of a fourth neutrino in section~\ref{sec:prob}, we confront it with the atmospheric neutrino data of AMANDA and IceCube-40 in section~\ref{sec:analysis}. Our conclusion will be given at section~\ref{sec:conc}.

\section{\label{sec:prob}Oscillation probability in 3+1 scheme}

The 3+1 scheme for neutrino masses consist of three mostly active neutrinos with masses $(m_1,m_2,m_3)$ that accommodate the observation of solar and atmospheric oscillations, and a mostly sterile state with mass $m_4$ separated from active states by $\Delta m_{41}^2\sim1 \,{\rm eV}^2 \gg \Delta m_{21,31}^2$. In this scheme, due to the large $\Delta m_{41}^2$ and small active-sterile mixing, the effect of the sterile neutrino on the solar neutrino oscillation and conventional atmospheric neutrino oscillation ($E_\nu \sim $~GeV) is negligible. However, the new large mass-squared difference induces an active-sterile oscillation at short base-lines $\sim10$~m for neutrinos with energy $E_\nu\sim100$~MeV, which is invoked to interpret the LSND, MiniBooNE, reactor and gallium anomalies~\cite{Peres:2000ic,Barger:1998bn,Grimus:2001mn,Maltoni:2001mt,Karagiorgi:2009nb,Kopp:2011qd,Giunti:2011gz,Giunti:2011hn,Giunti:2011cp}. 

The addition of one neutrino state leads to a generalization of the PMNS matrix to a $4\times4$ unitary matrix and introduces three new mass-squared differences $\Delta m_{4i}^2$ ($i=1,2,3$). However, since $\Delta m_{41}^2 \gg \Delta m_{21,31}^2$, the 3+1 model effectively introduces four new parameters to the oscillation phenomenology: one mass-squared difference $\Delta m_{41}^2$ and three angles $(\theta_{14},\theta_{24},\theta_{34})$ describing the active-sterile mixing. Assuming that all the CP-violating phases vanish, the $4\times4$ unitary mixing matrix $\mathbf{U_4}$ can be parametrized in the following way~\cite{deGouvea:2008nm}:
\begin{equation}\label{eq:para}
\mathbf{U_4}= \mathbf{R^{34}}(\theta_{34})\mathbf{R^{24}}(\theta_{24})\mathbf{R^{14}}(\theta_{14})\mathbf{R^{23}}(\theta_{23})\mathbf{R^{13}}(\theta_{13})\mathbf{R^{12}}(\theta_{12})~,
\end{equation}
where $\mathbf{R^{ij}}(\theta_{ij})$ ($i,j=1,\ldots,4$ and $i<j$) is the $4\times4$ rotation matrix in the $ij$-plane with the angle $\theta_{ij}$, with elements
\begin{equation}
\left[\mathbf{R^{ij}}(\theta_{ij})\right]_{kl}=(\delta_{ik}\delta_{il}+\delta_{jk}\delta_{jl})c_{ij}+(\delta_{ik}\delta_{jl}-\delta_{il}\delta_{jk})s_{ij}+\left[(1-\delta_{ik})(1-\delta_{jl})+(1-\delta_{il})(1-\delta_{jk})\right]\delta_{kl}~,
\end{equation}
where $c_{ij}\equiv\cos\theta_{ij}$ and $s_{ij}\equiv\sin\theta_{ij}$.

The atmospheric neutrinos propagate through the Earth before detection at up-going zenith angles at the South Pole. The evolution of the neutrino flavors inside the Earth is described by the following equation ($\alpha,\beta=e,\mu,\tau,s$)
\begin{equation}\label{eq:evolution}
i\frac{{\rm d}\nu_\alpha}{{\rm d}r}=\left[ \frac{1}{2E_\nu}\mathbf{U_4} \mathbf{M^2}\mathbf{U_4}^\dagger+\mathbf{V}(r) \right]_{\alpha\beta}\nu_\beta-\frac{i}{2}\mathbf{\Gamma}_{\alpha\alpha}\nu_\alpha~,
\end{equation}  
where $\mathbf{M^2}$ is the mass-squared differences matrix given by
\begin{equation}
\mathbf{M^2}=\mathbf{{\rm diag}}\left(0,\Delta m_{21}^2, \Delta m_{31}^2, \Delta m_{41}^2\right)~.
\end{equation}
The diagonal matrix $\mathbf{V}(r)$ is the matter potential as a function of distance $r$ given by
\begin{equation}
\mathbf{V}(r)=\sqrt{2}G_F\mathbf{{\rm diag}}\left(N_e(r), 0,0, N_n(r)/2\right)~.
\end{equation}
Here $N_e(r)$ and $N_n(r)$ are the electron and neutron number density of the Earth which we have fixed using the PREM model~\cite{prem}. The last term in Eq.~(\ref{eq:evolution}) takes into account the absorption of neutrinos inside the Earth; and the absorption matrix $\mathbf{\Gamma}$ is given by
\begin{equation}
\mathbf{\Gamma}={\rm diag} \left( \Gamma_e,\Gamma_\mu,\Gamma_\tau,0 \right)~,
\end{equation}
where $\Gamma_\alpha=\sigma_{\nu_\alpha p}N_p(r)+\sigma_{\nu_\alpha n}N_n(r)$. $N_p(r)$ and $N_n(r)$ are the proton and neutron number densities of the Earth and $\sigma_{\nu_\alpha n(p)}$ is the total (charged current + neutral current) interaction cross section of $\nu_\alpha$ with neutrons (protons). The neutrality of Earth implies that $N_e(r)=N_p(r)$ and, to a good approximation $N_n(r)=N_p(r)$: departures are less than 3\%. Also, since the absorption of neutrinos inside the Earth is important for $E_\nu \gtrsim 10$~TeV, the charged lepton mass effect in the charged current cross section is negligible and we can assume that $\Gamma_e=\Gamma_\mu=\Gamma_\tau\equiv\Gamma$. The same equations as Eq.~(\ref{eq:evolution}) describe the propagation of anti-neutrinos after replacing the cross section for neutrinos with those for anti-neutrinos and changing the sign of $\mathbf{V}(r)$.

The flux of atmospheric neutrinos with energies $E_\nu\gtrsim100$~GeV is mainly composed of $\nu_\mu$ and $\bar{\nu}_\mu$. The $\nu_e$ and $\bar{\nu}_e$ flux is more than one order of magnitude smaller. The $\nu_\tau(\bar{\nu}_\tau)$ flux is negligible up to energies of $E_\nu\sim 100$~TeV where the charm contribution may surpass the conventional atmospheric flux. The onset of this flux has not been observed. IceCube detects the Cherenkov radiation from muons produced in charged current interaction of $\nu_\mu$ and $\bar{\nu}_\mu$ with the nuclei inside or in the vicinity of detector. Thus, in our analysis we consider the survival probability of muon (anti-)neutrinos. The baseline of up-going atmospheric neutrinos detected by IceCube detector varies from zero for horizontal neutrinos to the diameter of the Earth $2R_\oplus\sim1.3\times10^4$~km for vertical up-going events. For neutrinos with the energy $E_\nu\gtrsim100$~GeV, the flavor oscillations resulting from the $\Delta m_{21}^2$ and $\Delta m_{31}^2$ mass splittings are negligible and the survival probability of muon (anti-)neutrinos in the 3+1 scheme in vacuum is given by  
\begin{equation}
P\big(\nu_\mu(\bar{\nu}_\mu)\to\nu_\mu(\bar{\nu}_\mu)\big)\simeq 1-\sin^2 2\theta_{\mu\mu} \sin^2\left(\frac{\Delta m_{41}^2L}{4E_\nu}\right)~,
\end{equation}
where $\sin^2 2\theta_{\mu\mu}\equiv4|U_{\mu4}|^2(1-|U_{\mu4}|^2)$. With the parametrization chosen in Eq.~(\ref{eq:para}) we have $U_{\mu4}=c_{14}s_{24}$. The short base-line appearance experiments MiniBooNE ($\nu_\mu\to\nu_e$ and $\bar{\nu}_\mu\to\bar{\nu}_e$) and LSND ($\bar{\nu}_\mu\to\bar{\nu}_e$) are sensitive to $\sin^2 2\theta_{e\mu}\equiv4|U_{e4}|^2|U_{\mu4}|^2$, while possible disappearance of the reactor $\bar{\nu}_e$ flux constrains the mixing parameter $\sin^2 2\theta_{ee}\equiv4|U_{e4}|^2(1-~|U_{e4}|^2)$. However, the three parameters ($\theta_{ee},\theta_{e\mu},\theta_{\mu\mu}$) are not independent and the limits on two of them can be translated to the other one; see Refs.~\cite{Kopp:2011qd,Giunti:2011gz,Giunti:2011hn} for the global analysis and the interdependency of the mixing parameters. 

Although the vacuum survival probability for muon (anti-)neutrinos only depends on the $\theta_{14}$ and $\theta_{24}$, because of the critical role of matter effects through MSW resonant flavor conversion in our analysis, the survival probability also depends on the $\theta_{34}$. This can be seen by writing the evolution equation Eq.~(\ref{eq:evolution}) in the so-called propagation basis where the total Hamiltonian is diagonal. To illustrate the point, consider the evolution equation written in the basis $(\nu_e,\nu^\prime_\mu,\nu^\prime_\tau,\nu^\prime_s)^T=\mathbf{R^{34}}\mathbf{R^{24}}(\nu_e,\nu_\mu,\nu_\tau,\nu_s)^T$. It is easy to check that the presence of the matter potential $\mathbf{V}$ and its nonzero element $N_n$, through the term $\mathbf{R^{34}}\cdot\mathbf{R^{24}}\cdot\mathbf{V}\cdot\mathbf{R^{24\dagger}}\cdot\mathbf{R^{34\dagger}}$, results in the dependence of the survival probability on $\theta_{34}$ (for the details see Ref.~\cite{Razzaque:2011ab}). Also, taking into account the current upper limit on the $\sin^22\theta_{ee}\equiv\sin^22\theta_{14}$ from the reactor disappearance experiments, the dependence of the survival probability on $\theta_{14}$ is very weak. We therefore make the approximation that $\theta_{14}=0$, which implies $\theta_{\mu\mu}\equiv\theta_{24}$.

\begin{figure}[ht!]
\begin{center}
\subfloat[$P(\bar{\nu}_\mu\to\bar{\nu}_\mu)$ for $\sin^22\theta_{\mu\mu}=0.001$]{
 \includegraphics[width=0.5\textwidth]{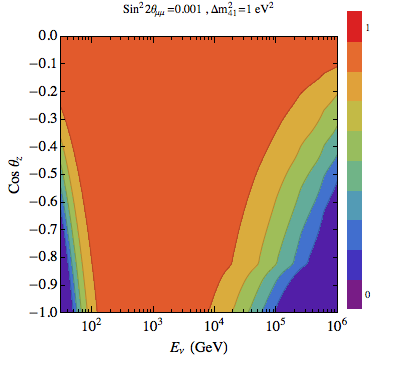}
  \label{fig:mubar001}
}
\subfloat[Subfigure 2 list of figures text][$P(\bar{\nu}_\mu\to\bar{\nu}_\mu)$ for $\sin^22\theta_{\mu\mu}=0.01$]{
 \includegraphics[width=0.5\textwidth]{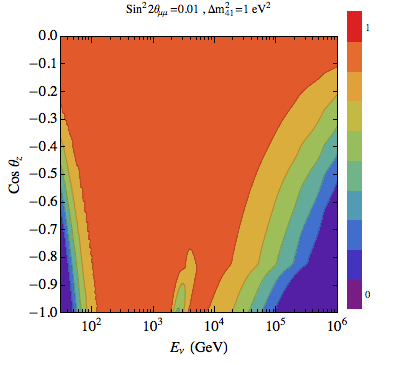}
 \label{fig:mubar01}
}
\qquad
\subfloat[Subfigure 3 list of figures text][$P(\bar{\nu}_\mu\to\bar{\nu}_\mu)$ for $\sin^22\theta_{\mu\mu}=0.1$]{
  \includegraphics[width=0.5\textwidth]{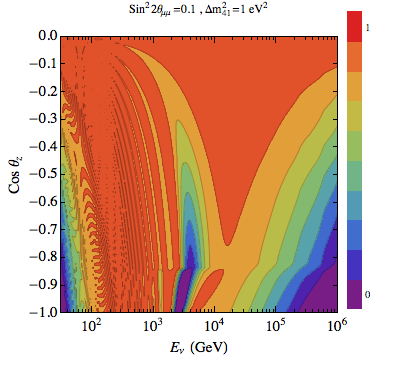}
  \label{fig:mubar1}
}
\subfloat[Subfigure 4 list of figures text][$P(\bar{\nu}_\mu\to\bar{\nu}_\mu)$ for $\sin^22\theta_{\mu\mu}=0.3$]{
   \includegraphics[width=0.5\textwidth]{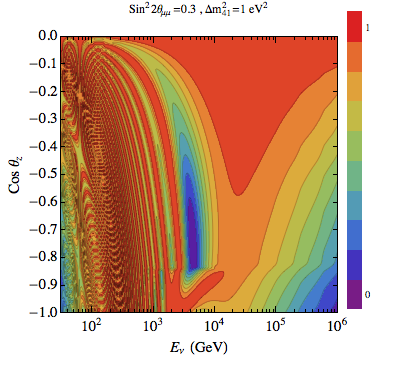}
    \label{fig:mubar3}
}
\end{center}
\caption{\label{fig:probmubar}The oscillogram for the survival probability $P(\bar{\nu}_\mu\to\bar{\nu}_\mu)$ for $\Delta m_{41}^2=1\,{\rm eV}^2$, $\sin^2\theta_{34}=0$ and the different values of $\sin^22\theta_{\mu\mu}$ indicated in each sub-caption. The survival probability includes the attenuation of the neutrino flux inside the Earth from the charged and neutral current interaction with nuclei.}
\end{figure}

\begin{figure}[ht!]
\begin{center}
\subfloat[$P(\nu_\mu\to\nu_\mu)$ for $\sin^22\theta_{\mu\mu}=0.001$]{
 \includegraphics[width=0.5\textwidth]{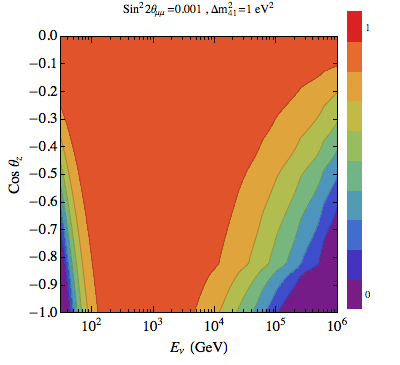}
  \label{fig:mu001}
}
\subfloat[Subfigure 2 list of figures text][$P(\nu_\mu\to\nu_\mu)$ for $\sin^22\theta_{\mu\mu}=0.01$]{
 \includegraphics[width=0.5\textwidth]{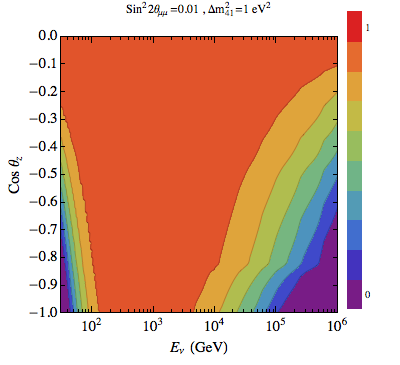}
 \label{fig:mu01}
}
\qquad
\subfloat[Subfigure 3 list of figures text][$P(\nu_\mu\to\nu_\mu)$ for $\sin^22\theta_{\mu\mu}=0.1$]{
  \includegraphics[width=0.5\textwidth]{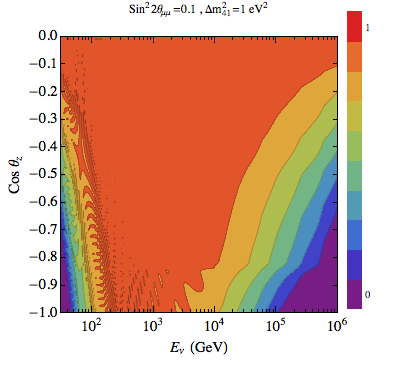}
  \label{fig:mu1}
}
\subfloat[Subfigure 4 list of figures text][$P(\nu_\mu\to\nu_\mu)$ for $\sin^22\theta_{\mu\mu}=0.3$]{
   \includegraphics[width=0.5\textwidth]{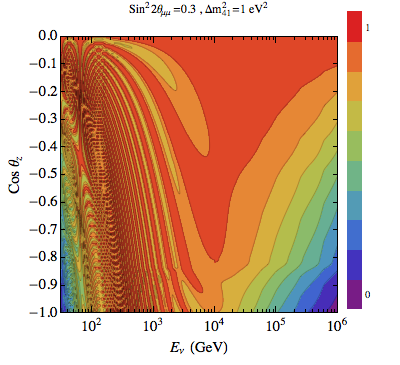}
    \label{fig:mu3}
}
\end{center}
\caption{\label{fig:probmu}The same as Fig~\ref{fig:probmubar} but for $P(\nu_\mu\to\nu_\mu)$.}
\end{figure}

Fig.~\ref{fig:probmubar} shows the oscillogram of the $\bar{\nu}_\mu$ survival probability for different values $\sin^22\theta_{\mu\mu}$ and for fixed value $\Delta m_{41}^2=1\,{\rm eV}^2$ and $\theta_{34}=0$, taking into account the absorption inside the Earth. In each oscillogram the axes are neutrino energy and the cosine of the zenith angle at IceCube. With increasing values of $\sin^22\theta_{\mu\mu}$ the presence of a dip in the survival probability develops for neutrino energies $\sim$~TeV. The decrease in the survival probability at higher energies and vertical directions ($\cos\theta_z\simeq -1$) comes from the attenuation of the neutrino flux by the Earth (the last term in Eq.~(\ref{eq:evolution})). Also, for near vertical directions ($\cos\theta_z\sim[-1,-0.8]$) the parametric resonance resulting from the propagation of neutrinos through the alternating mantle-core-mantle densities inside the Earth, will play a role~\cite{Akhmedov:1998ui}. To compare the survival probabilities for $\bar{\nu}_\mu$ and $\nu_\mu$, in Fig.~\ref{fig:probmu} we show the oscillograms of $P(\nu_\mu\to\nu_\mu)$ for the same mixing parameters as Fig.~\ref{fig:probmubar}. For the $\nu_\mu$ survival probability there is no MSW effect and therefore no dip. The absence of the MSW resonance for $\nu_\mu$ is a consequence of the normal hierarchy that we assumed between the active and sterile neutrino masses ($\Delta m_{41}^2>0$). For the inverted hierarchy ($\Delta m_{41}^2<0$) the resonance flavor conversion occurs for the $\nu_\mu$ channel and not for $\bar{\nu}_\mu$. However, from cosmological considerations, the $\Delta m_{41}^2<0$ is strongly disfavored, because, in this case, the three active neutrino masses will all be at the $\sim$~eV scale. In our analysis we consider the normal case ($\Delta m_{41}^2>0$) only.

\section{\label{sec:analysis}Parameter constraints from IceCube-40 and AMANDA-II data}

The number of events expected in IceCube can be calculated by the convolution of the neutrino flux with the effective area $A_{\rm eff}$ of the detector:
\begin{equation}\label{eq:number}
N=T(2\pi)\left[\int A_{\rm eff}^\nu (E_\nu,\cos\theta_z) \Phi_\nu (E_\nu,\cos\theta_z) \,{\rm d}E_\nu \,{\rm d}\cos\theta_z+ (\nu\to\bar{\nu})\right]~,
\end{equation} 
where $T$ is the livetime of data-taking and the $2\pi$ factor comes from the integration over the azimuthal angle. The $\Phi_{\nu(\bar{\nu})} (E_\nu,\cos\theta_z)$ is the flux of atmospheric (anti-)neutrinos as a function of neutrino energy and zenith angle in units ${\rm GeV}^{-1}\,{\rm m}^{-2}\,{\rm s}^{-1}\,{\rm sr}^{-1}$. The conventional atmospheric neutrino flux comes from the decay of pions and kaons produced in the interactions of cosmic rays with the Earth's atmosphere. From a few GeV to ~$\sim100$~TeV this flux dominates and we will parametrize it following its calculation by~\cite{Honda:2006qj}. For energies $\gtrsim100$~TeV the contribution from the decay of charmed mesons and baryons may play a role, the so-called prompt flux, and we use the calculation of~\cite{Enberg:2008te} to represent its possible contribution in this analysis.

\begin{figure}[ht!]
\begin{center}
\subfloat[]{
 \includegraphics[width=0.5\textwidth]{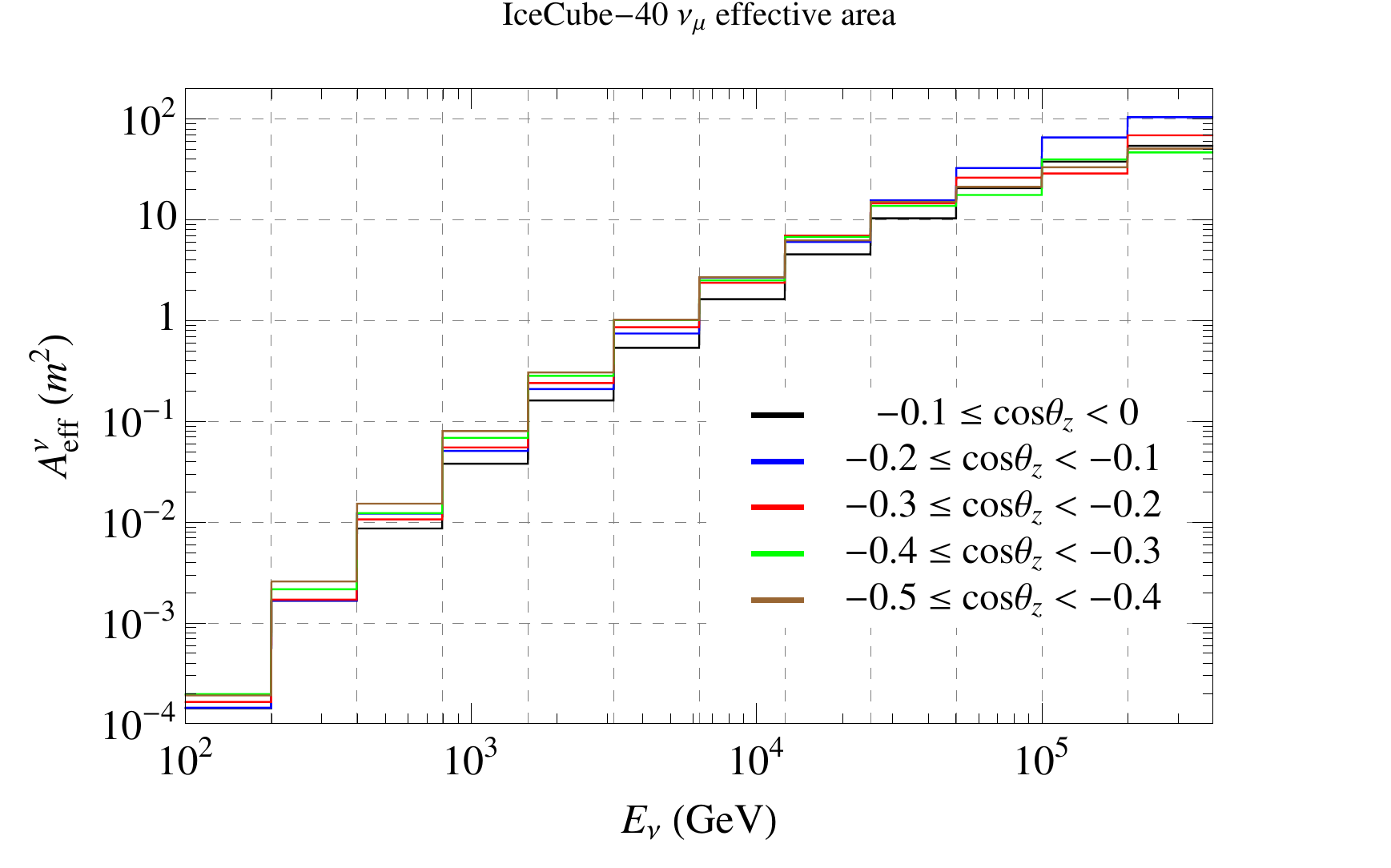}
  \label{fig:effic40-1}
}
\subfloat[]{
 \includegraphics[width=0.5\textwidth]{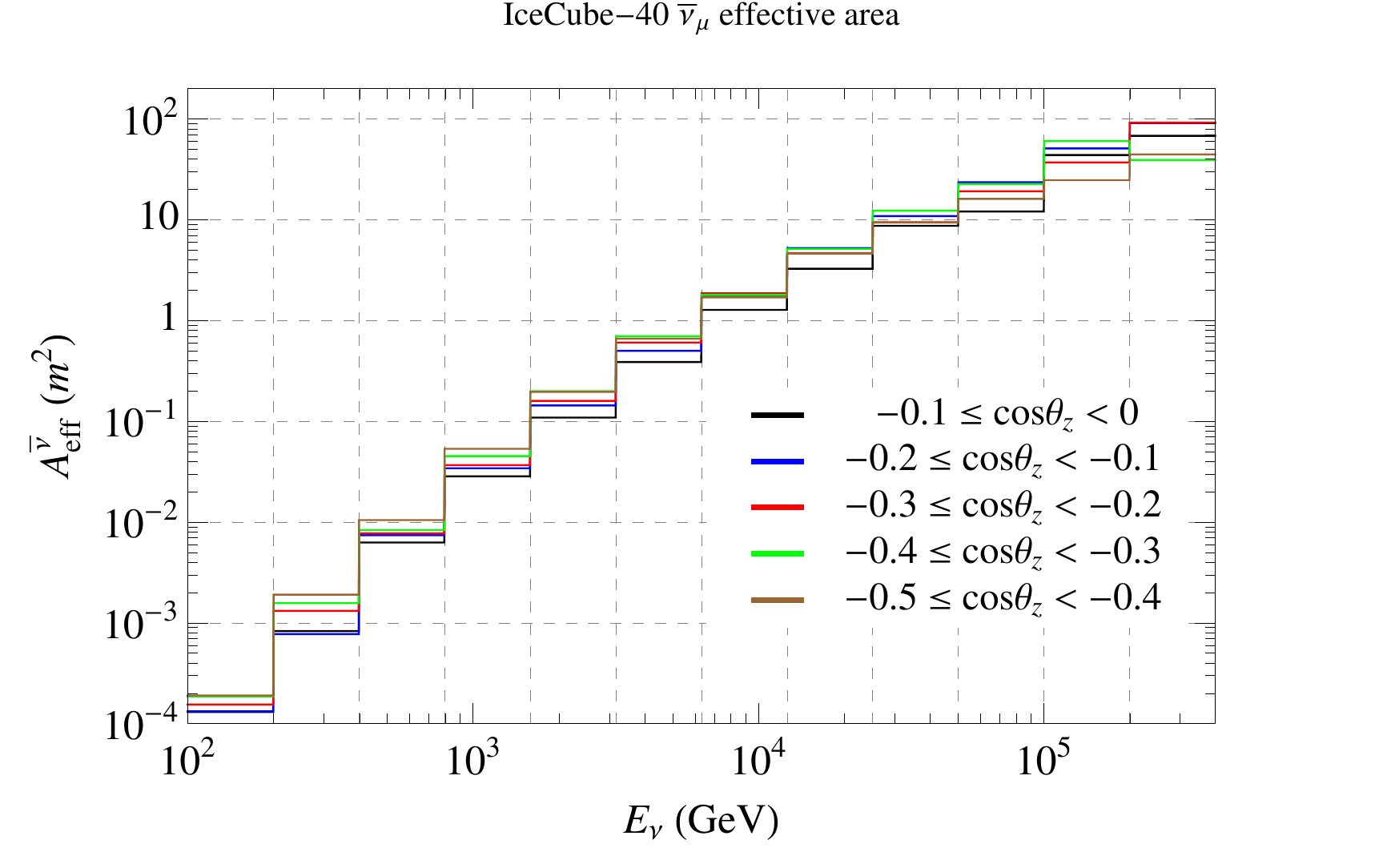}
 \label{fig:effic40-1bar}
}
\qquad
\subfloat[]{
  \includegraphics[width=0.5\textwidth]{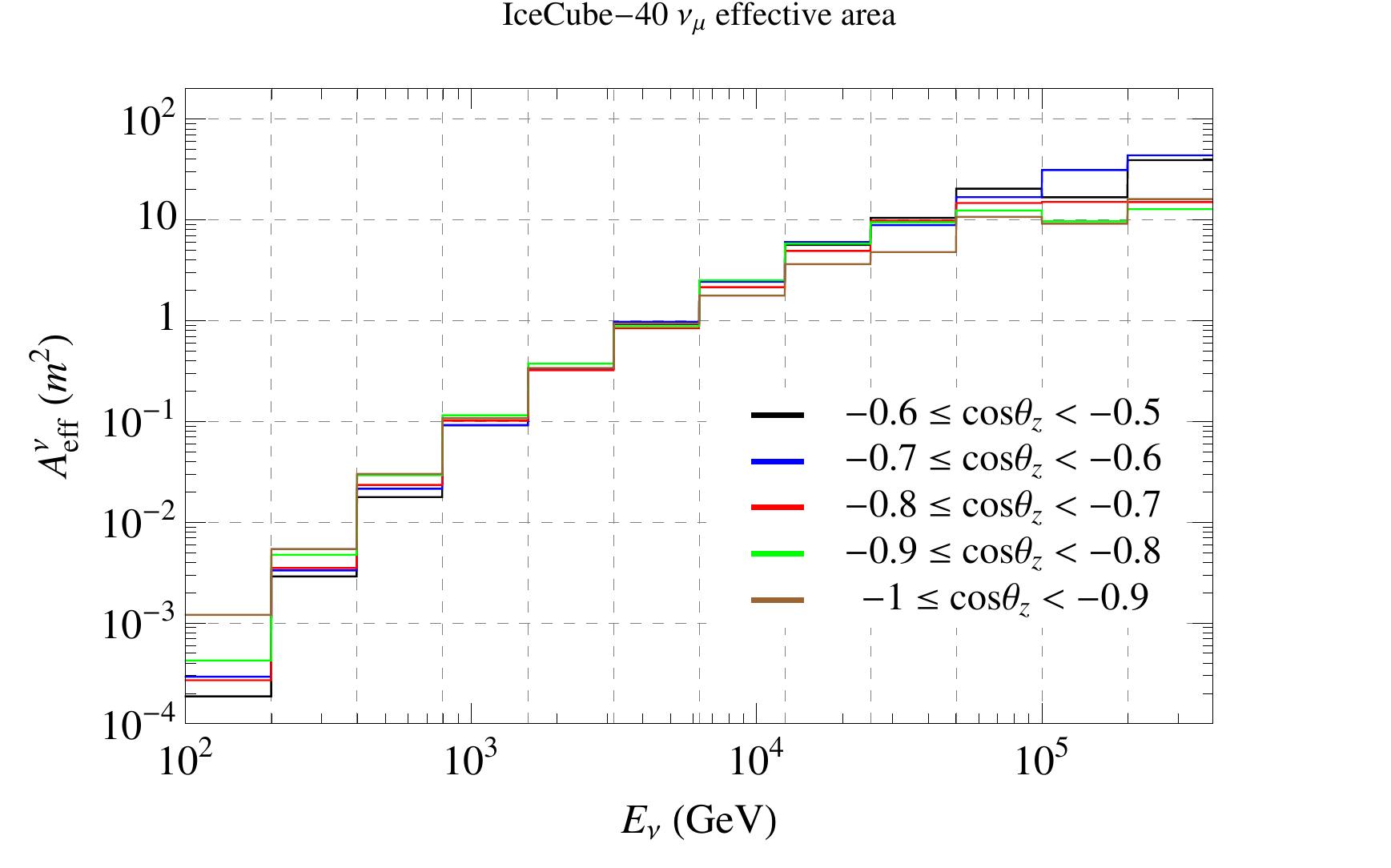}
  \label{fig:effic40-2}
}
\subfloat[]{
   \includegraphics[width=0.5\textwidth]{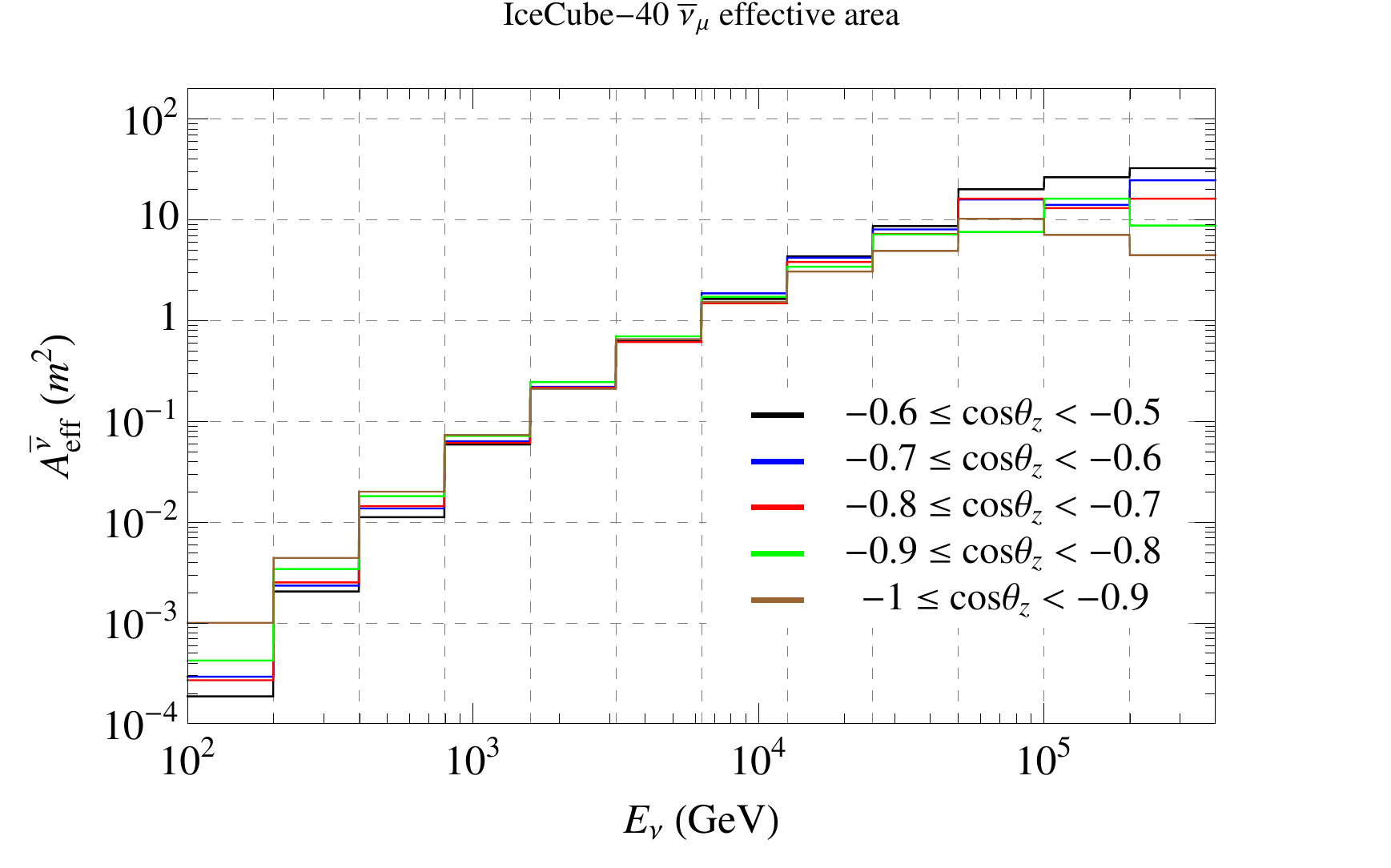}
    \label{fig:effic40-2bar}
}
\end{center}
\caption{\label{fig:effic40}The $A_{\rm eff}^\nu$ and $A_{\rm eff}^{\bar{\nu}}$ for IceCube-40 experiment as a function of neutrino energy and zenith angle. The vertical dashed grid lines show the binning in energy. The figures correspond to: (a) neutrino effective area for $-0.5\leq\cos\theta_z<0$; (b) anti-neutrino effective area for $-0.5\leq\cos\theta_z<0$; (c) neutrino effective area for $-1\leq\cos\theta_z<-0.5$; (d) anti-neutrino effective area for $-1\leq\cos\theta_z<-0.5$~.}
\end{figure}

\begin{figure}[ht!]
\begin{center}
\subfloat[]{
 \includegraphics[width=0.5\textwidth]{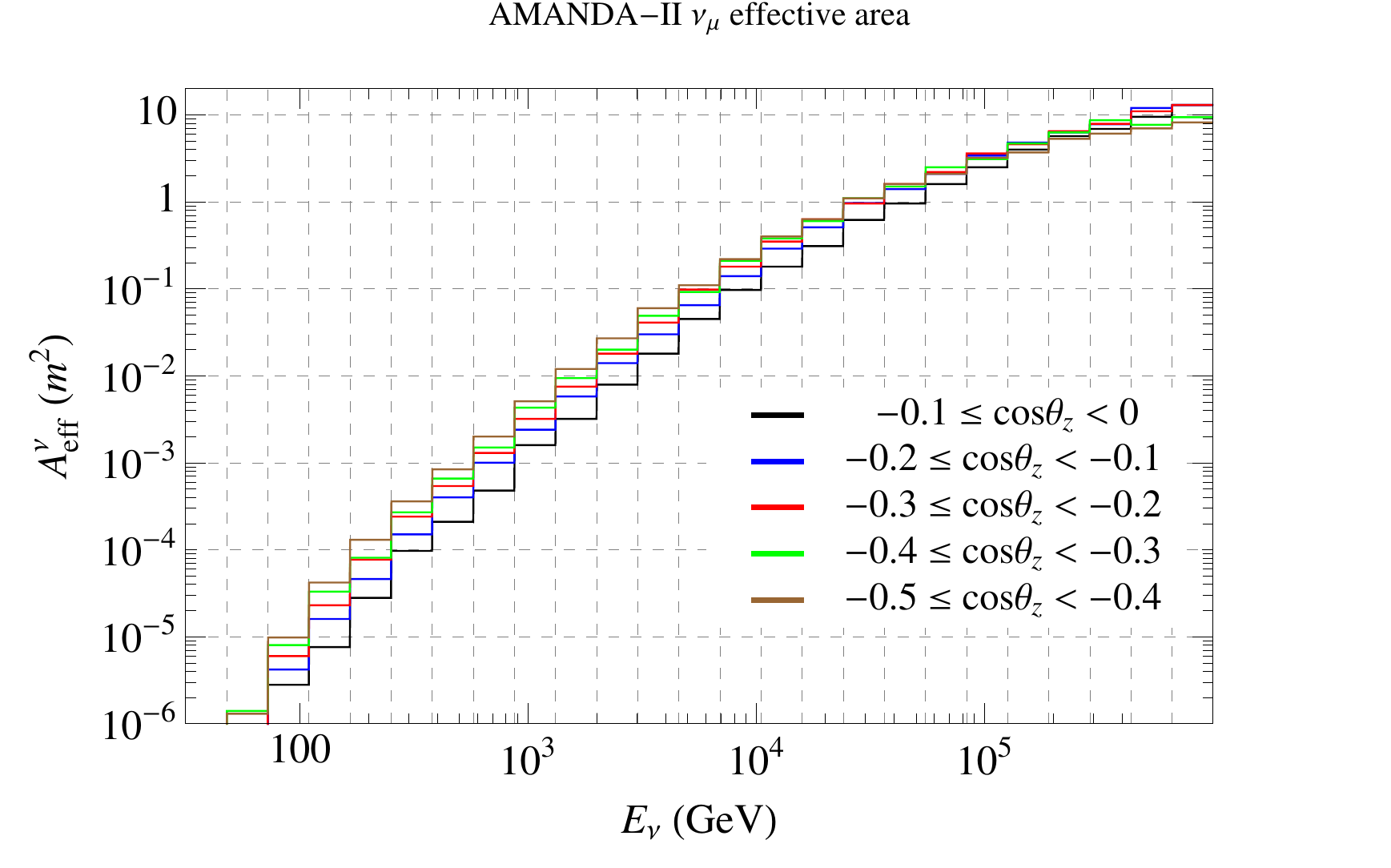}
  \label{fig:effamanda-1}
}
\subfloat[]{
 \includegraphics[width=0.5\textwidth]{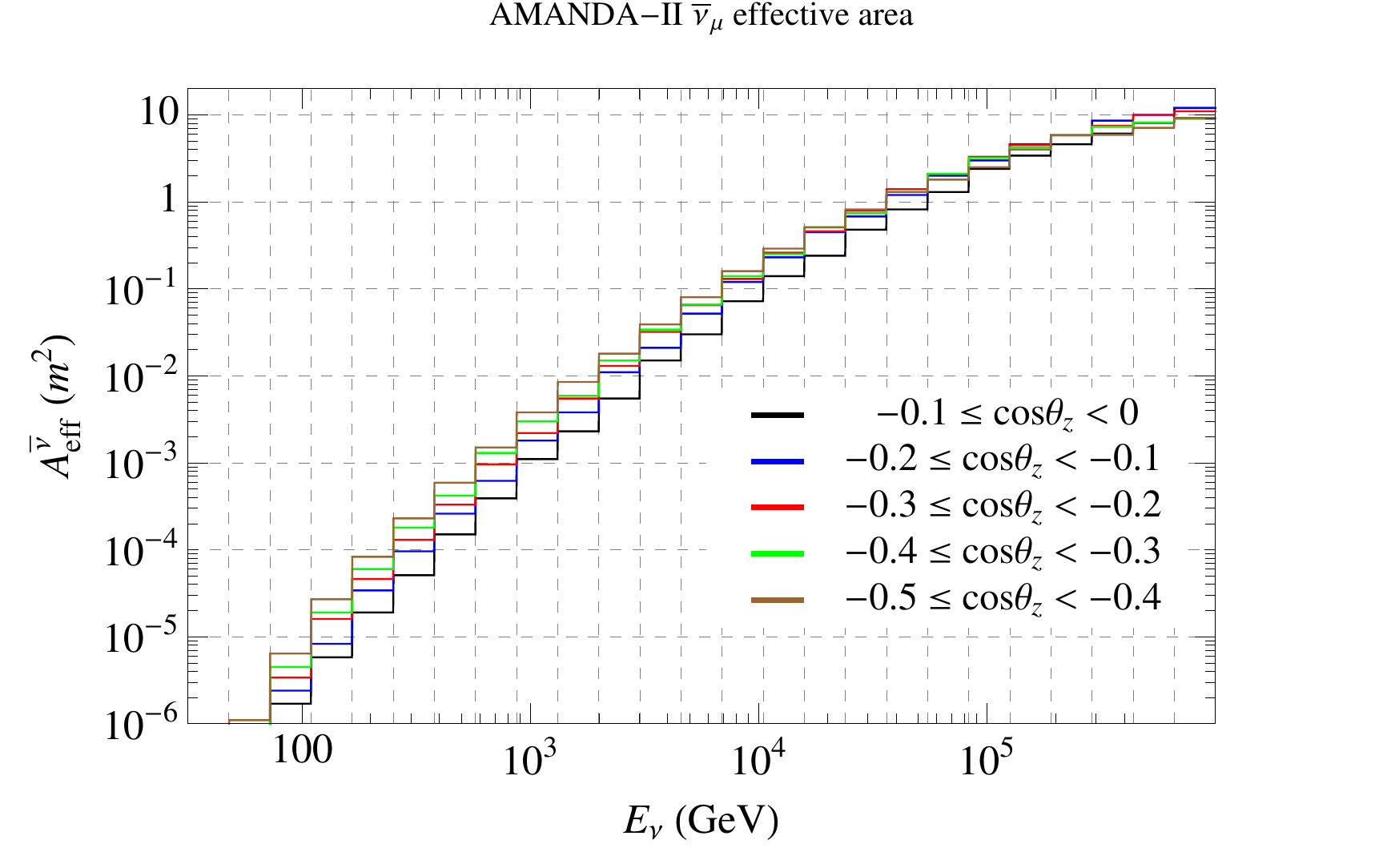}
 \label{fig:effamanda-1bar}
}
\qquad
\subfloat[]{
  \includegraphics[width=0.5\textwidth]{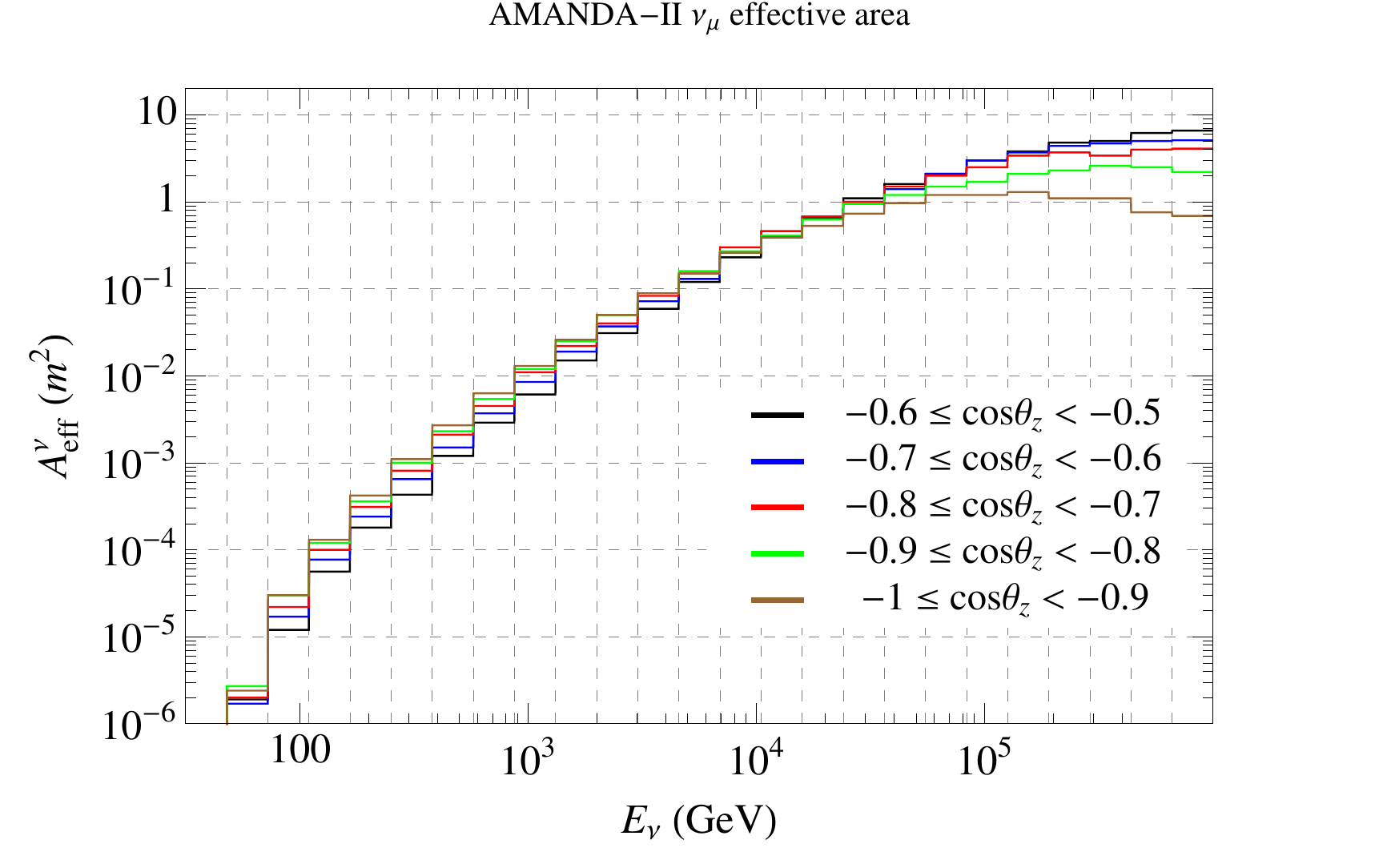}
  \label{fig:effamanda-2}
}
\subfloat[]{
   \includegraphics[width=0.5\textwidth]{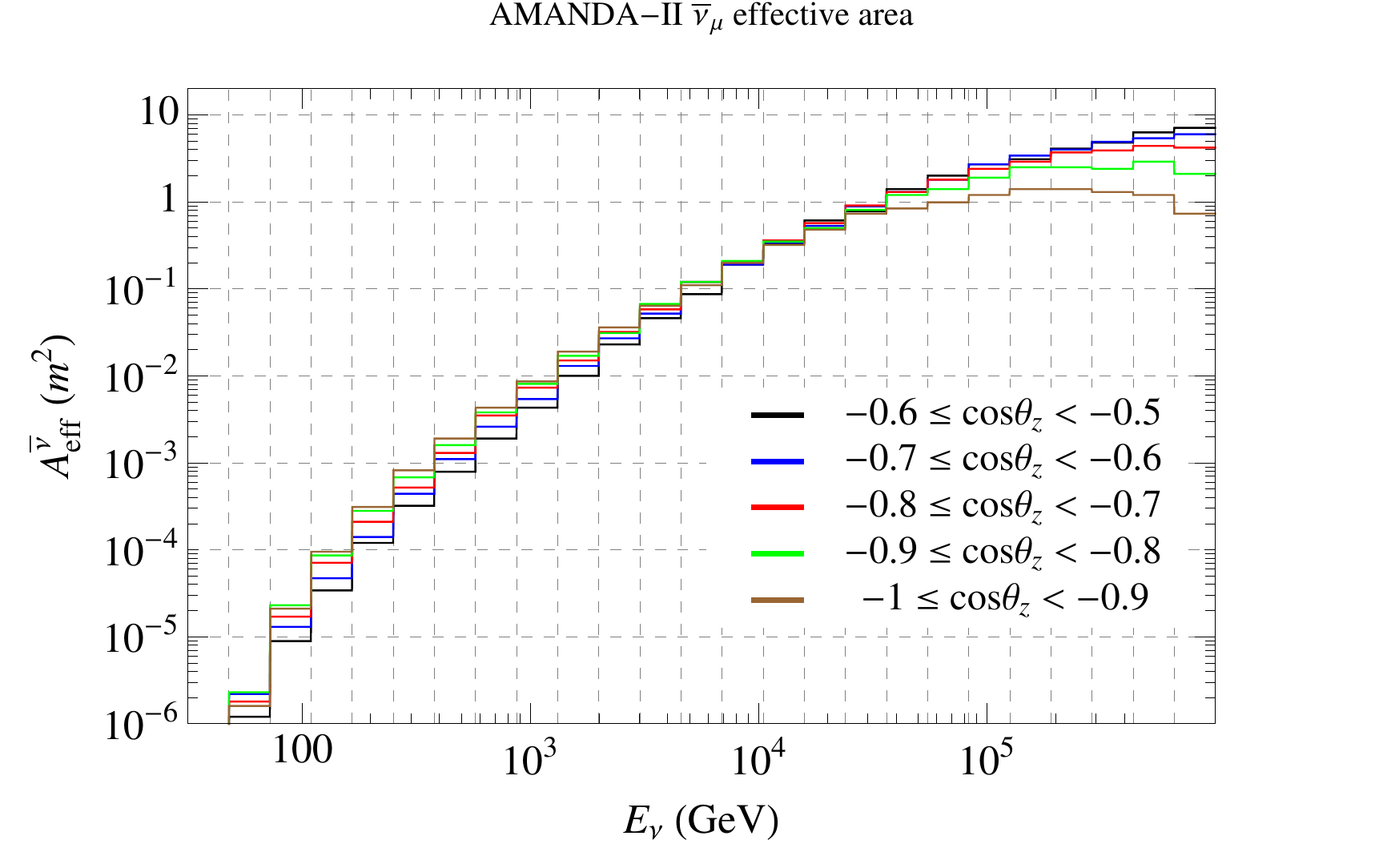}
    \label{fig:effamanda-2bar}
}
\end{center}
\caption{\label{fig:effamanda}The same as Fig.~\ref{fig:effic40} but for AMANDA-II experiment~\cite{kelley}.}
\end{figure}

We analyze the data of the IceCube-40 and AMANDA-II experiments. The IceCube-40 experiment measured the atmospheric neutrino flux in the energy range 100 GeV$-$400 TeV with a livetime of $T=359$~days; we use the effective area (separately for neutrinos and anti-neutrinos) in ten bins of $\cos\theta_z$ (with bin width $0.1$) and 12 bins of $E_\nu$ (with bin width 0.3 in $\log_{10}(E_\nu/{\rm GeV})$)~\cite{Abbasi:2010ie}. The AMANDA-II experiment measured the atmospheric neutrino flux in the energy range $10^{1.5}-10^6$~GeV for 1387~days of data-taking~\cite{Abbasi:2008ih,amandaweb}. We used the effective area (separately for neutrinos and anti-neutrinos) in ten bins of $\cos\theta_z$ (with bin width $0.1$) and 25 bins of $E_\nu$ (with bin width 0.18 in $\log_{10}(E_\nu/{\rm GeV})$)~\cite{kelley}. The complete set of effective areas we used in our analysis are shown in Figs.~\ref{fig:effic40}~and~\ref{fig:effamanda} at the end of paper.

In the calculation of the number of events using Eq.~(\ref{eq:number}), the effective area is convoluted with the atmospheric flux of neutrinos at the surface of Earth. This means that the attenuation of the neutrino flux inside the Earth is encoded in the effective area. But, in the presence of sterile neutrinos at 3+1 model, the attenuation of the neutrino flux is different from the standard $3\nu$ framework due to the singlet nature of sterile neutrinos. In the $3\nu$ framework and for energies $E_\nu\gtrsim100$~GeV the $\mathbf{\Gamma}$ term in Eq.~(\ref{eq:evolution}) is flavor blind (diagonal) which enables us to factor out the attenuation term from the evolution equation. In this case the attenuation can be incorporated into the probability by a multiplicative exponential factor $\exp[\Gamma X(\cos\theta_z)]$, where $X(\cos\theta_z)$ is the slant depth as a function of zenith angle. In order to take into account the difference in the absorption  of neutrinos inside the Earth between $3\nu$ and 3+1 models, we divide the effective area in each bin by the averaged attenuation exponential factor for $3\nu$ model in that bin. Then, in the calculation of number of events in Eq.~(\ref{eq:evolution}), we insert the probability of oscillation including the absorption for the 3+1 model.  

To illustrate the reach of the IceCube-40 and AMANDA data to constrain the 3+1 model, we define the following $\chi^2$ function
\begin{equation}
\chi^2 (\Delta m^2_{41},\theta_{34},\theta_{24};\alpha) = \sum_i \frac{\left(N^{\rm data}_{i}-\alpha N^{3+1}_{i}(\Delta m^2_{41},\theta_{34},\theta_{24})\right)^2}{\sigma_i^2}+\frac{(1-\alpha)^2}{\sigma_{\alpha}^2}~,
\end{equation}
where $N^{\rm data}_{i}$ is the observed number of events in $i^{\rm th}$ bin of $\cos\theta_z$ and $N^{3+1}_{i}$ is the expected number of events in the $i^{\rm th}$ bin of $\cos\theta_z$ assuming the 3+1 model with the mixing parameters ($\Delta m^2_{41},\theta_{34},\theta_{24}$). The $\sigma_i=\sqrt{N^{\rm data}_{i}}$ in the denominator represents the statistical error in the observed events. The factor $\alpha$ is the normalization factor that represents the uncertainty in the normalization of the atmospheric neutrino flux with $\sigma_\alpha=0.3$~.

\begin{figure}[t]
\centering
\includegraphics[scale=0.67]{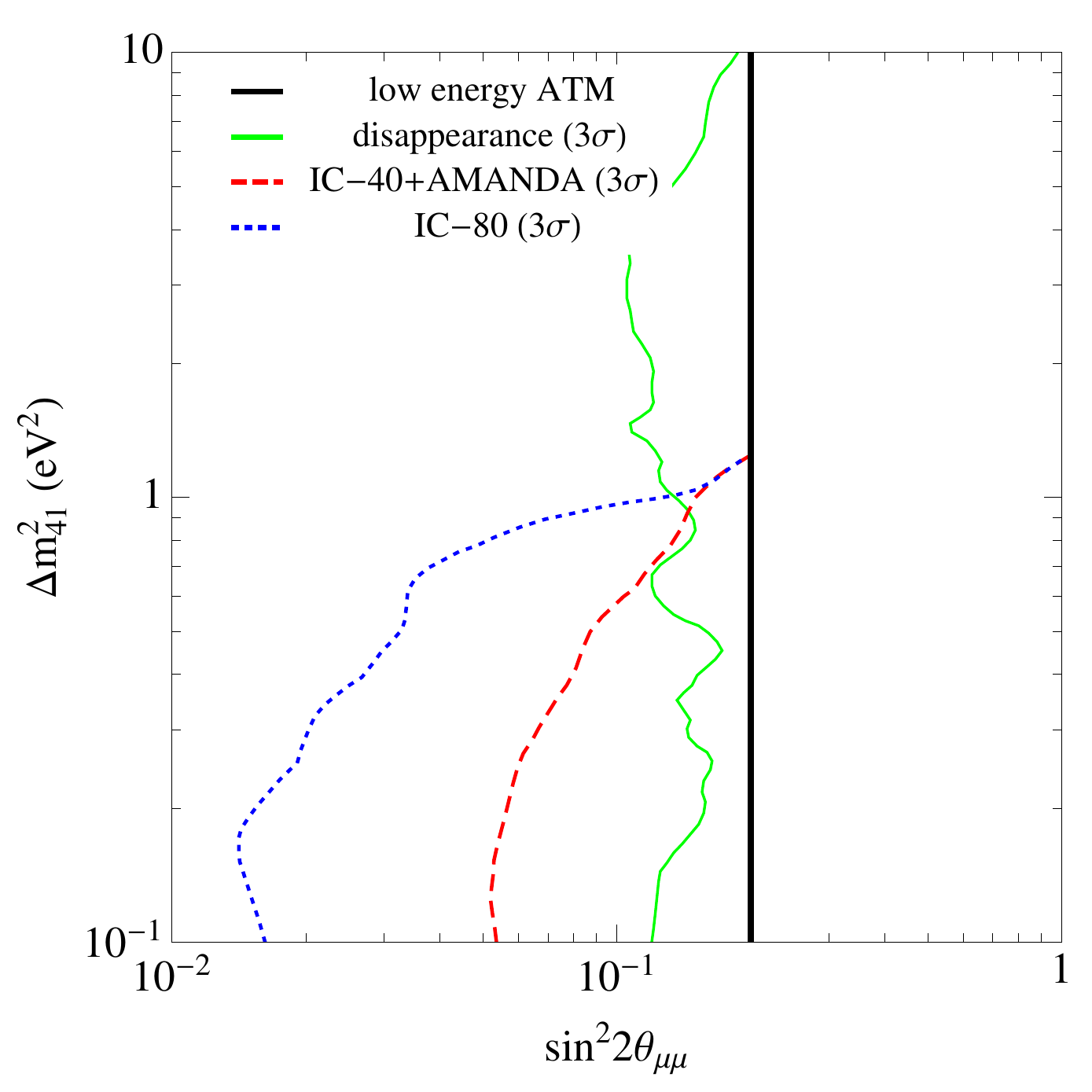}
\caption{\label{fig:limit} The allowed region in the $(\Delta m_{41}^2,\sin^22\theta_{\mu\mu})$ parameter space from the combined analysis IceCube-40 and AMANDA-II data (red dashed curve) at $3\sigma$ confidence level. The green solid curve shows the $3\sigma$ upper limit from the disappearance data taken from Ref.~\cite{Giunti:2011hn}. The blue dotted curve shows an estimate of the $3\sigma$ sensitivity of IceCube-80 after three years of data-taking. The black vertical line shows the upper limit from low energy atmospheric neutrino oscillation data~\cite{Maltoni:2007zf}.}
\end{figure}

Fig.~\ref{fig:limit} shows the $3\sigma$ confidence level allowed region in the $(\Delta m_{41}^2,\sin^22\theta_{\mu\mu})$ plane from the combined analysis IceCube-40 and AMANDA-II data (red dashed curve). The allowed region corresponding to the IceCube-40 data set alone, is very close to red curve in Fig.~\ref{fig:limit}, which means that due to the lower statistics, the contribution of AMANDA-II data set to constraining the parameter space is small. The best-fit to the data corresponds to the $3\nu$ framework. Introducing a sterile neutrino deteriorates the fit (increases the $\chi^2$ value). The black solid vertical line shows the upper limit on the mixing angle (independent from the mass-splitting value) which comes from the non-observation of the oscillation pattern in the low energy down-going atmospheric neutrinos which leads to requirement of large mixing elements corresponding to light mass states  ({\it i. e.} large $\sum_{i=1}^{3}|U_{\mu i}|^2$)~\cite{Maltoni:2007zf}. The green solid curve shows the upper limit from the disappearance data at $3\sigma$ confidence level (including reactor anomaly, MINOS and CDHSW)~\cite{Giunti:2011hn}. In the range $\Delta m_{41}^2\sim 0.1-1~{\rm eV}^2$ relevant for the interpretation of the recent experimental anomalies, the combined IceCube and AMANDA data strongly constrain the parameter space. The blue dotted curve shows the prospect for the complete IceCube detector to constrain the 3+1 model after three years of data-taking. However, it should be mentioned that for the sensitivity of IceCube-80, we have used a coarse binned non-optimized effective area~\cite{GonzalezGarcia:2009jc,danninger} and the sensitivity of the full IceCube80+DeepCore should be significantly better. The weaker constraint for larger values of $\Delta m_{41}^2$ results from the fact that by increasing the value of the mass-splitting the dip in the survival probability shifts to higher energies; where the flux of atmospheric neutrinos is reduced and the statistics worse. This will be remedied by future data from the completed detector. For the $\Delta m_{41}^2\sim 0.1-1~{\rm eV}^2$ the resonance occurs at $0.3-3~{\rm TeV}$, where the flux of atmospheric neutrinos is sizable and the detection efficiency is optimized.

According to the analysis performed in this letter, the $3+1$ scheme is disfavored. From the global fit of short base-line experiments data and the MiniBooNE neutrino and anti-neutrino data\footnote{The analysis of~\cite{Giunti:2011cp} performed before the latest MiniBooNE data release~\cite{AguilarArevalo:2012va}, which the excess of events observed in both neutrino and anti-neutrino channels. In~\cite{Giunti:2011cp}, the data set of~\cite{AguilarArevalo:2008rc} has been considered.}, the best-fit values of the $3+1$ mixing parameters are $(\Delta m_{41}^2,\sin^22\theta_{\mu\mu})=(0.9~{\rm eV}^2,0.083)$~\cite{Giunti:2011cp}. The data of AMANDA-II and IceCube-40 exclude this point at $2\sigma$ level. After three years of data-taking with IceCube-80 it is possible exclude this point at $\sim 3.2\sigma$ level.

The Refs.~\cite{Razzaque:2011ab}~and~\cite{Barger:2011rc} also look for the evidence of sterile neutrino in IceCube-40 data. In~\cite{Razzaque:2011ab} the authors show that the $3+1$ scheme is disfavored in comparison with the $3\nu$ case. However in the analysis of~\cite{Razzaque:2011ab}, the same effective area for neutrinos and anti-neutrinos has been used, that can mislead the result. Also, the parameter space of $3+1$ is not scanned in~\cite{Razzaque:2011ab} and they calculate the $\chi^2$ value for a few points in the $0.5~{\rm eV}^2 < \Delta m^2_{41} < 3.0~{\rm eV}^2$. In the Ref.~\cite{Barger:2011rc} also, the authors consider the IceCube-40 data with the cut $E_\nu > 332$~GeV and again the effective area used in the analysis is not optimized. In~\cite{Barger:2011rc} the authors conclude that more information is necessary to draw a conclusion about the sterile neutrino hypotheses. In this letter we used the full AMANDA-II and IceCube-40 data sets with the widest energy distribution available. We considered fine-binned effective neutrino area separately for neutrinos and anti-neutrinos. Also, we implemented a full scan of the parameter space : $0.1~{\rm eV}^2 < \Delta m^2_{41} < 10~{\rm eV}^2$ for mass-squared difference and complete range for the mixing parameter $\sin^22\theta_{\mu\mu}$.

\section{\label{sec:conc}Summary and Conclusion}

The presence of a fourth sterile neutrino state leads to a distortion of the zenith angle distribution of high energy atmospheric neutrinos through the MSW active-sterile resonance inside the Earth. The energy range of the resonance is ideally covered by the South Pole detectors, IceCube and AMANDA. The presence of this resonance is not supported by the present evidence. We have shown that the 3+1 model is severely constrained. The limit obtained in our analysis is stronger than the combined limit of all other experiments, for the active-sterile mass-squared differences $\Delta m^2_{41}\lesssim 1\,{\rm eV}^2$. Also, we have shown that the completed IceCube detector can test the 3+1 model for mixing angle as small as $\sin^22\theta_{\mu\mu}\sim10^{-2}$. The complete detector has already collected data for more than one year and IceCube will therefore further constrain the presence of sterile neutrino, or possibly discover it.

\begin{acknowledgments} 
The work of F.~H. is supported in part by the National Science Foundation under Grant No. OPP-0236449, by the DOE under grant DE-FG02-95ER40896 and in part by the University of Wisconsin Alumni Research Foundation. O.~L.~G.~P. thanks the ICTP, FAPESP, CAPES and Fulbright commission for financial support, the C.N. Yang Institute at Stony Brook University and Arizona State University for the hospitality were this work was partially developed. A.~E. thanks FAPESP and World Premier International Research Center Initiative (WPI Initiative), MEXT, Japan, for financial support. The authors thank CENAPAD and CCJDR for computing facilities.
\end{acknowledgments}

\newpage

\newpage

\end{document}